\documentclass[preprint,12pt,authoryear]{elsarticle}

\usepackage[T1]{fontenc}
\usepackage{lmodern}
\usepackage{amsmath,amssymb}
\usepackage{graphicx}
\usepackage{booktabs}
\usepackage{array}
\usepackage{siunitx}
\usepackage{xcolor}
\usepackage{xurl}
\usepackage{hyperref}
\newcommand{\dataset}{AstroClassifier}
\newcommand{\balacc}{balanced accuracy}
\newcommand{\macrof}{macro-F1}
\journal{Astronomy and Computing}

\begin{document}
\begin{frontmatter}

\title{Object-independent evaluation of small-data astronomical image classifiers:
architecture, representation, and acquisition shortcuts}

\author[unesp]{Eraldo Pereira Marinho\corref{cor1}}
\ead{pereira.marinho@unesp.br}
\cortext[cor1]{Corresponding author}
\affiliation[unesp]{organization={Institute of Geoscience and Exact Sciences,
S\~ao Paulo State University (UNESP)},
addressline={Av. 24 A, 1515},
city={Rio Claro},
postcode={13506-900},
state={S\~ao Paulo},
country={Brazil}}

\begin{abstract}
Reported performance in small astronomical image datasets can reflect object reuse,
field-of-view differences, acquisition settings, and catalogue-derived properties as well as astronomical
morphology. We designed \dataset{} to measure these effects in a four-class task
comprising galaxies, globular clusters, nebulae, and open clusters. The dataset
contains 370 catalogue-supported physical objects, each represented by three
Digitized Sky Survey FITS bands, and uses a group-constrained split of 222 training,
75 validation, and 73 test objects. We compared a custom CNN--Transformer trained
from scratch with ImageNet-pretrained ResNet-50 and ViT-B/16 under an adaptive
full-field representation, a common angular field, and two central-region-masked
controls, using three fixed seeds. In the one-pass object-independent test, mean
balanced accuracy on the adaptive view was $0.773\pm0.014$, $0.874\pm0.054$, and
$0.866\pm0.023$ for the custom model, ResNet-50, and ViT-B/16, respectively. No
primary architecture contrast rejected its null after Holm correction; ResNet-50
and ViT-B/16 differed by 0.007 balanced-accuracy points (95\% cluster-bootstrap
interval $[-0.048,0.062]$). Controls based on acquisition and catalogue metadata
reached 0.721--0.735, and central-region-masked adaptive views reached 0.731--0.800.
These results demonstrate generalization to unseen physical objects, but also show
that image context, acquisition settings, and catalogue-derived object properties
carry substantial class information. In this regime, high accuracy
cannot be interpreted as evidence that a classifier learned target morphology, and
the comparison does not isolate architecture from pretraining.
\end{abstract}


\end{frontmatter}

\section{Introduction}

Deep learning is widely used for astronomical image analysis, from galaxy morphology
to survey triage~\cite{dieleman2015,huertas2018,baron2019machine}. Its apparent
success can be difficult to interpret when datasets contain many files but few
independent astronomical objects. Multiple bands, renders, crops, or observations of
the same target do not create independent physical examples. Related products that
cross data splits can yield high scores without demonstrating generalization to an
unseen object. Class-dependent acquisition choices, including field of view (FOV),
pixel scale, and target extent, can also become shortcuts
\cite{geirhos2020shortcut,kapoor2023leakage}.

The AstroClassifier project began by asking whether a hybrid convolutional neural
network (CNN) and Transformer could classify astronomical images with few labeled
examples. An audit of the original experiment showed that a credible answer required
a different statistical design. The central question became: how much class
information transfers to previously unseen physical objects when object identity,
representation, and acquisition cues are explicitly examined?

We make four contributions:
\begin{enumerate}
\item a catalogue-supported dataset in which the physical object, rather than a band
image or derivative, is the observational unit;
\item a prospectively specified Dataset-v2 comparison of a custom CNN--Transformer,
ResNet-50, and ViT-B/16,
with configuration selection restricted to development data;
\item matched common-field, central-region-masked, and metadata-only controls that expose
class information outside the nominal target morphology; and
\item a one-pass, label-isolated test procedure that records predictions before an
independent scoring process accesses the labels.
\end{enumerate}

This manuscript reports the prospectively specified Dataset-v2 evaluation on 73
previously unseen objects. The model families reflect the earlier exploratory history
of AstroClassifier; dataset construction, split assignment, model-selection rules,
controls, and held-out analyses were specified before test evaluation. The
label-isolated results separate evidence of object-independent prediction from the
stronger claim that the models learned target morphology.

\section{Related work}

CNNs exploit local spatial structure and have produced strong results for
astronomical morphology~\cite{dieleman2015,huertas2018}. Residual connections enable
deeper convolutional models~\cite{he2016deep}. Transformers replace convolutional
locality with content-dependent attention~\cite{vaswani2017}; ViT applies attention
to image patches and benefits strongly from large-scale pretraining
\cite{dosovitskiy2020,khan2021}.

Small-data comparisons must distinguish architectural capacity from pretraining,
augmentation, and model selection. Augmentation can improve generalization
\cite{shorten2019}, but derivatives of one object remain statistically dependent.
High validation accuracy may also arise from shortcuts that remain stable within a
collection but fail under acquisition or identity shifts~\cite{geirhos2020shortcut}.
Leakage and inappropriate split units are recognized threats to reproducibility in
machine-learning-based science~\cite{kapoor2023leakage}. We therefore treat the
architecture comparison as one component of a group-aware evaluation.

\section{Dataset and experimental design}

\subsection{Objects, provenance, and split}

The candidate list was assembled prospectively after excluding identities recovered
from the legacy experiment. It contained 400 primary candidates (100 per class) and
73 reserves. Quality assessment of the primary cohort produced 370 eligible objects,
13 rejected product bundles, and 17 deferred cases; among the reserves, 66 remained
unused and seven were ineligible under deterministic checks. Galaxies span broad
morphological groupings; the nebula class includes planetary, HII, reflection,
dark, and supernova-remnant candidates; open clusters use catalogue open-cluster
identifications; and globular clusters draw on the Milky Way globular-cluster
catalogue (December 2010 revision)~\cite{harris1996}. Object identities and aliases
were reconciled against catalogue records,
including SIMBAD~\cite{wenger2000simbad}, before splitting. No legacy image,
prediction, or model outcome entered candidate selection.

The resulting eligible pool contains 370 physical objects:
98 galaxies, 92 globular clusters, 85 nebulae, and 95 open clusters. Each has one
observation bundle containing DSS2 blue, red, and near-infrared FITS products
retrieved through the CDS HiPS2FITS service~\cite{fernique2015,boch2020hips2fits}.
The pool therefore contains 1,110 FITS
files but only 370 distinct astronomical targets. Bands, previews, quality-control
renders, and deterministic derivatives inherit the identity of their parent object.
Table~\ref{tab:composition} reports the class-dependent FOV distribution. Its large
differences, especially for nebulae and open clusters, motivate the common-field and
metadata controls rather than being treated as incidental preprocessing details.

\begin{table*}[t]
\centering
\caption{Composition of the eligible pool and five-number summary of source
FOV in arcmin. Values are computed per physical object before splitting.}
\label{tab:composition}
\resizebox{\textwidth}{!}{%
\begin{tabular}{lrrrrrr}
\toprule
Class & Objects & Minimum & First quartile & Median & Third quartile & Maximum \\
\midrule
Galaxies         & 98 & 5 & 5 & 5 & 5 & 23 \\
Globular clusters& 92 & 5 & 5 & 5 & 6.14 & 15.64 \\
Nebulae          & 85 & 5 & 5 & 15 & 15 & 120 \\
Open clusters    & 95 & 5 & 15 & 27.2 & 71.5 & 120 \\
\bottomrule
\end{tabular}
}
\end{table*}

Quality eligibility was fixed before model training. Same-object relations,
derivative ownership, and overlapping-field components were converted into
indivisible assignment units. Seventeen non-trivial components were preserved. The
allocation is shown in Table~\ref{tab:split}; test contains 73 objects in 61 units.

\begin{table}[t]
\centering
\caption{Final object counts by class and split.}
\label{tab:split}
\begin{tabular}{lrrrrr}
\toprule
Split & Galaxies & Globular & Nebulae & Open & Total \\
\midrule
Training   & 59 & 55 & 51 & 57 & 222 \\
Validation & 20 & 19 & 17 & 19 & 75 \\
Test       & 19 & 18 & 17 & 19 & 73 \\
\bottomrule
\end{tabular}
\end{table}

\subsection{Representations and preprocessing}

All representations use verified FITS bytes and blue/red/NIR channel order. For each
band, first and 99th percentiles are computed on the prescribed view. Values are
clipped to $[0,1]$, transformed as
\begin{equation}
u=\frac{\operatorname{asinh}(10z)}{\operatorname{asinh}(10)},
\end{equation}
and mapped to $[-1,1]$ by $(u-0.5)/0.5$. Stored tensors are contiguous
\texttt{float32} arrays with shape $3\times224\times224$. The channels are not
conventional RGB; using them with ImageNet models is a fixed transfer policy rather
than an assertion of spectral equivalence.

Four deterministic variants isolate different information:
\begin{description}
\item[Adaptive full field (A).] The original $512\times512$ field, spanning 5 to 120
arcmin, is retained and resized by exact area-overlap averaging. It preserves context
and class-dependent FOV and scale.
\item[Common field (B).] Each target is resampled to a 4-arcmin tangent-plane field
with 15 arcsec pixels and 15 arcsec Gaussian FWHM. The effective $16\times16$ grid is
expanded by repeating every cell $14\times14$.
\item[Central-region-masked controls.] In $A_{\mathrm{outer}}$, the central
$128\times128$ region is set to zero; in $B_{\mathrm{outer}}$, the central
$4\times4$ cells are zeroed. Stretch percentiles use unmasked pixels only. Diffuse
emission and neighbours may remain, so these are context controls rather than
certified blank sky.
\end{description}

\begin{figure}[p]
\centering
\includegraphics[width=\textwidth]{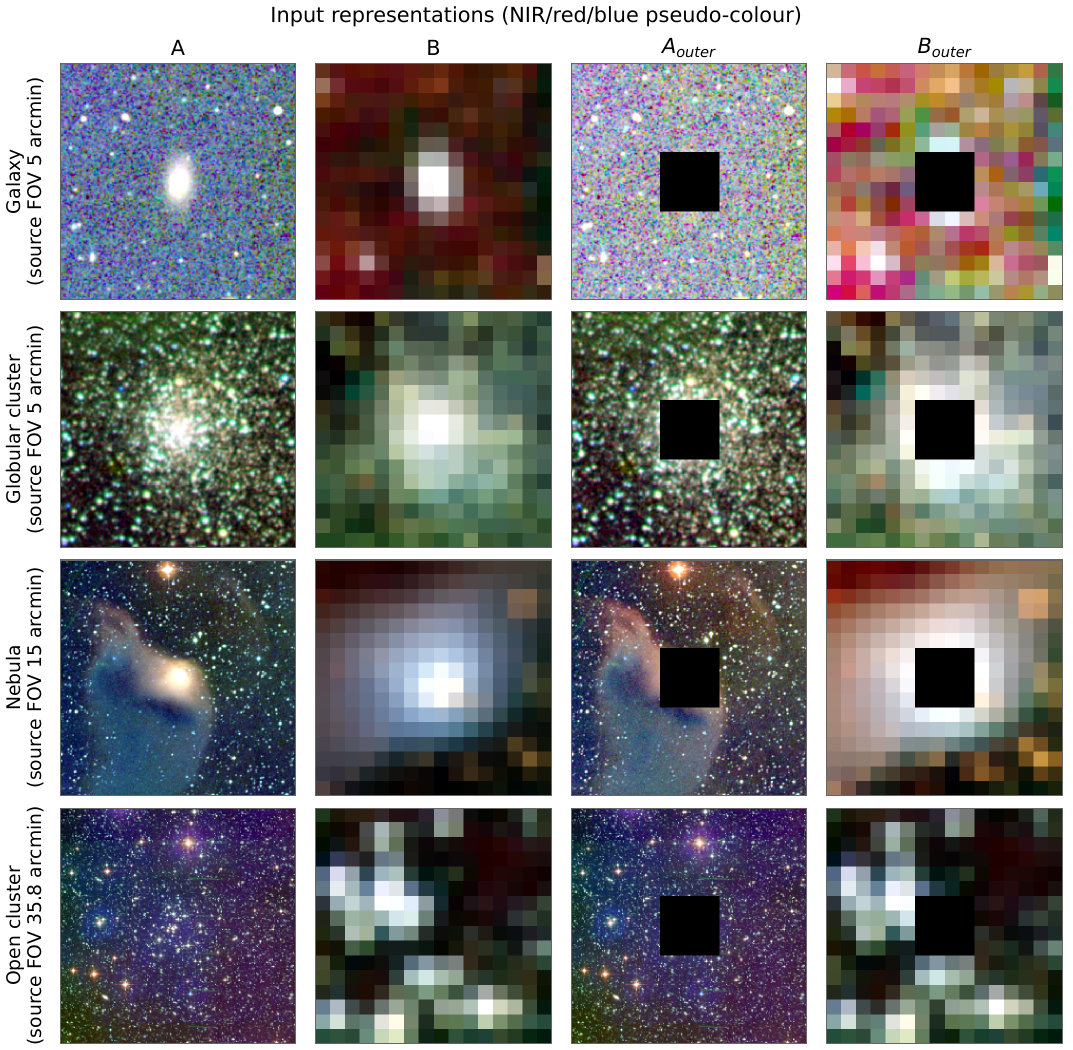}
\caption{Examples of the four input representations. Rows show one
deterministically selected validation object from each class; columns show A, B,
$A_{\mathrm{outer}}$, and $B_{\mathrm{outer}}$. Stored blue/red/NIR channels are
displayed as NIR/red/blue pseudo-colour for visualization only. Black regions are
the prescribed central masks. The source FOV shown at left applies to A and
$A_{\mathrm{outer}}$; B uses the common 4-arcmin field for every object.}
\label{fig:representations}
\end{figure}

Figure~\ref{fig:representations} shows one validation object per class under all
four transformations. Within each class, the example is the object closest
to the validation-set median source FOV, with object identifier used to break ties.
The visible block structure in B and $B_{\mathrm{outer}}$ is the prescribed
$16\times16$ effective sampling, not a rendering artifact.

For every training object and epoch, one of the eight elements of the dihedral group
$D_4$ is selected deterministically from the physical-object identifier and epoch
number. The transformation can therefore change across epochs while remaining
exactly reproducible; the same rotation/reflection is applied to all channels.
Validation and test use the identity transform. No random crops, colour jitter,
test-time augmentation, or family-specific normalization is applied.

\subsection{Acquisition and catalogue metadata controls}

Class membership explains 0.5827 of the descriptive variance of log(FOV) in the
curated pool. Two fixed controls use no pixels: multinomial logistic regression with
balanced class weights and a depth-3 decision tree. Inputs are log source FOV, log
source pixel scale, log catalogue extent, extent missingness, and an FOV-cap
indicator. Imputation and scaling are fit on training objects only. Names, catalogue
types, aliases, paths, and indices are excluded. These controls measure available
non-image signal. FOV, pixel scale, and the cap indicator describe acquisition,
whereas catalogue extent and its missingness describe the target. The controls
therefore measure their combined predictive information; they are not
acquisition-only controls and do not by themselves prove causal leakage.

\section{Models and training}

\subsection{Model families}

The custom model is trained from scratch. Four convolutional stages with 128, 256,
512, and 1,024 channels produce a feature map. Fixed mean-plus-gradient pooling
reduces it to an $8\times8$ grid: 64 tokens with 128-dimensional embeddings. Fixed
separable two-dimensional sinusoidal positions precede two Transformer encoder
layers with eight attention heads. Mean token pooling and a linear head yield four
logits. A deterministic bilinear adapter maps the stored $224\times224$ input to the
selected model's $256\times256$ contract without changing the representation files.

The reference CNN is torchvision ResNet-50~\cite{he2016deep}; the reference
Transformer is torchvision ViT-B/16~\cite{dosovitskiy2020}. Both use locally
preserved official ImageNet-1K weights fixed before training.
Their heads are replaced by four-output linear layers.

\begin{table*}[t]
\centering
\small
\caption{Training settings fixed before test evaluation. Every family uses batch
size 16, at most 100 epochs,
class-weighted cross-entropy, gradient-norm cap 3, ReduceLROnPlateau (factor 0.1,
patience 5), and early stopping with patience 20.}
\label{tab:models}
\resizebox{\textwidth}{!}{%
\begin{tabular}{@{}lllll@{}}
\toprule
Family & Initialization & Optimizer & Base LR & Weight decay \\
\midrule
Custom CNN--Transformer & From scratch & Adam  & $9\times10^{-5}$ & $5\times10^{-7}$ \\
ResNet-50               & ImageNet-1K   & AdamW & $1\times10^{-4}$ & $1\times10^{-4}$ \\
ViT-B/16                & ImageNet-1K   & AdamW & $3\times10^{-5}$ & $5\times10^{-2}$ \\
\bottomrule
\end{tabular}
}
\end{table*}

For pretrained models, the new head uses ten times the backbone learning rate and all
layers are trainable from the first epoch. This gives established models their
practical transfer-learning advantage; initialization remains part of the comparison.

\subsection{Pre-specified Dataset-v2 selection}

The representation-A search used only 222 training and 75 validation objects. The
candidate grid contained six custom, 12 ResNet-50, and 12 ViT-B/16 configurations. Each
was trained with seeds 3908274, 3908275, and 3908276, producing 90 fits. One
configuration per family was selected by mean validation \balacc{}, then mean
\macrof{}, mean weighted loss, sample variance of \balacc{}, and configuration ID.
All three seeds were retained.

Each winner was then applied unchanged to B, $A_{\mathrm{outer}}$, and
$B_{\mathrm{outer}}$, producing 27 control fits. Including the two metadata controls,
development used 119 fits: 117 neural and two metadata-only. Controls could not
trigger selection, retuning, or exclusion.

The families themselves were informed by exploratory work that preceded Dataset-v2;
the prospective claim applies to this Dataset-v2 selection and held-out evaluation,
not to the entire history of architecture development. Because hyperparameters were
selected on A and transferred without retuning, the other representations assess the
robustness of the A-selected training configurations. Their results do not estimate
the best performance attainable after representation-specific optimization.

\section{Evaluation and statistical analysis}

Primary metrics are \balacc{}, \macrof{}, and accuracy, supplemented by per-class
metrics and confusion matrices. Neural results are reported per seed and as mean
$\pm$ sample SD. Seeds describe training variability and are not independent test
objects.

Test point metrics use 73 objects; inference resamples keep all 61 assignment units
intact. The pre-specified analysis uses 20,000 paired cluster-bootstrap replicates and
20,000 paired unit-level randomization replicates from NumPy PCG64 seed 20261021.
The same draws are shared across models, representations, metrics, and seeds, with
Holm corrections in pre-specified comparison families.

Final evaluation comprised 38 model evaluations: three families, four representations,
and three seeds (36 neural checkpoints), plus two metadata models. A label-isolated
inference procedure recorded each set of predictions once before an independent
scoring process accessed the labels. No test result triggered retraining, replacement,
reselection, or repeated inference.

\section{Development-stage results}

\subsection{Primary-view selection}

Table~\ref{tab:valA} reports validation results of each selected configuration. These
numbers are development evidence, not final test performance. ResNet-50 and ViT-B/16
had similar means; ViT-B/16 varied more across seeds. The custom model trailed both
pretrained references.

\begin{table}[t]
\centering
\caption{Validation results on A; mean $\pm$ sample SD over three seeds.}
\label{tab:valA}
\begin{tabular}{lcc}
\toprule
Family & Balanced accuracy & Macro-F1 \\
\midrule
Custom CNN--Transformer & $0.764\pm0.016$ & 0.752 \\
ResNet-50               & $0.860\pm0.005$ & 0.859 \\
ViT-B/16                & $0.858\pm0.045$ & 0.856 \\
\bottomrule
\end{tabular}
\end{table}

\subsection{Representation and outer-field controls}

Changing from A to the common field B reduced mean validation performance for every
family. Masked views retained substantial signal: $A_{\mathrm{outer}}$ \balacc{}
ranged from 0.712 to 0.771 and $B_{\mathrm{outer}}$ from 0.586 to 0.700. These
controls do not establish that morphology is unused. They show that context and
acquisition-correlated information are too strong to interpret an architecture score
in isolation.

\begin{table*}[t]
\centering
\caption{Validation controls; mean $\pm$ sample SD over three fixed seeds.}
\label{tab:controls}
\resizebox{\textwidth}{!}{%
\begin{tabular}{llcc}
\toprule
Family & Representation & Balanced accuracy & Macro-F1 \\
\midrule
Custom CNN--Transformer & B                    & $0.688\pm0.020$ & $0.676\pm0.020$ \\
                        & $A_{\mathrm{outer}}$ & $0.728\pm0.024$ & $0.707\pm0.039$ \\
                        & $B_{\mathrm{outer}}$ & $0.586\pm0.029$ & $0.575\pm0.033$ \\
\addlinespace
ResNet-50               & B                    & $0.720\pm0.046$ & $0.719\pm0.046$ \\
                        & $A_{\mathrm{outer}}$ & $0.712\pm0.107$ & $0.703\pm0.113$ \\
                        & $B_{\mathrm{outer}}$ & $0.660\pm0.081$ & $0.659\pm0.079$ \\
\addlinespace
ViT-B/16                & B                    & $0.771\pm0.037$ & $0.766\pm0.038$ \\
                        & $A_{\mathrm{outer}}$ & $0.771\pm0.019$ & $0.769\pm0.020$ \\
                        & $B_{\mathrm{outer}}$ & $0.700\pm0.038$ & $0.700\pm0.038$ \\
\bottomrule
\end{tabular}
}
\end{table*}

Metadata-only logistic regression obtained validation \balacc{} 0.711, \macrof{}
0.709, and accuracy 0.720. The fixed tree obtained 0.713, 0.714, and 0.720. These
results corroborate the class--FOV association and justify retaining both controls.

\section{Object-independent test results}
\label{sec:testresults}

The final held-out evaluation produced 73 predictions for each of the 38 model
evaluations. Predictions were recorded before scoring, and the completeness check
found no missing, duplicate, or extra evaluation. Table~\ref{tab:test} reports the
neural results. All models performed well above four-class chance on A, showing
transfer to physical objects absent from training and validation.

\begin{table*}[t]
\centering
\small
\caption{Object-independent test results; mean $\pm$ sample SD over three fixed
seeds. A and B are the adaptive and common-field views; ``outer'' denotes the
corresponding central-mask control.}
\label{tab:test}
\resizebox{\textwidth}{!}{%
\begin{tabular}{llccc}
\toprule
Family & Representation & Balanced accuracy & Macro-F1 & Accuracy \\
\midrule
Custom CNN--Transformer & A                    & $0.773\pm0.014$ & $0.748\pm0.003$ & $0.785\pm0.016$ \\
                        & B                    & $0.704\pm0.050$ & $0.698\pm0.047$ & $0.699\pm0.049$ \\
                        & $A_{\mathrm{outer}}$ & $0.731\pm0.019$ & $0.694\pm0.045$ & $0.744\pm0.016$ \\
                        & $B_{\mathrm{outer}}$ & $0.625\pm0.016$ & $0.624\pm0.014$ & $0.626\pm0.016$ \\
\addlinespace
ResNet-50               & A                    & $0.874\pm0.054$ & $0.873\pm0.056$ & $0.877\pm0.055$ \\
                        & B                    & $0.737\pm0.030$ & $0.733\pm0.029$ & $0.735\pm0.029$ \\
                        & $A_{\mathrm{outer}}$ & $0.737\pm0.036$ & $0.733\pm0.032$ & $0.740\pm0.041$ \\
                        & $B_{\mathrm{outer}}$ & $0.608\pm0.014$ & $0.605\pm0.021$ & $0.607\pm0.016$ \\
\addlinespace
ViT-B/16                & A                    & $0.866\pm0.023$ & $0.866\pm0.023$ & $0.868\pm0.021$ \\
                        & B                    & $0.793\pm0.015$ & $0.787\pm0.014$ & $0.790\pm0.016$ \\
                        & $A_{\mathrm{outer}}$ & $0.800\pm0.041$ & $0.799\pm0.041$ & $0.804\pm0.042$ \\
                        & $B_{\mathrm{outer}}$ & $0.667\pm0.017$ & $0.664\pm0.015$ & $0.662\pm0.016$ \\
\bottomrule
\end{tabular}
}
\end{table*}

\begin{figure}[htbp]
\centering
\includegraphics[width=\textwidth]{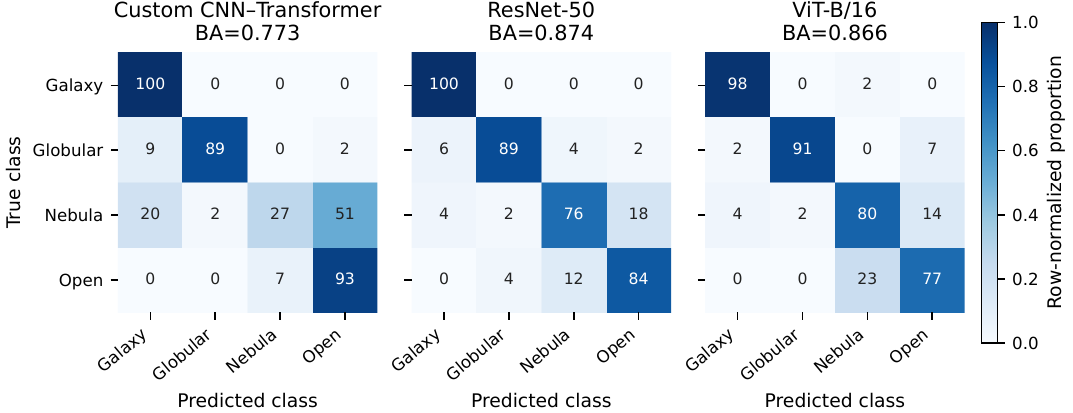}
\caption{Row-normalized confusion matrices for representation A. Each cell is the
mean over the three fixed seeds and is labelled as a percentage; rows are true
classes and columns are predictions. BA denotes mean balanced accuracy. Rounded
entries may not sum to exactly 100.}
\label{fig:confusionA}
\end{figure}

Figure~\ref{fig:confusionA} localizes the largest family difference. The custom
model classified only 27.5\% of nebulae correctly on average and assigned 51.0\%
of them to the open-cluster class. The corresponding nebula recalls were 76.5\%
for ResNet-50 and 80.4\% for ViT-B/16. Other class-specific trade-offs remain: for
example, ViT-B/16 more often assigned open clusters to the nebula class.

\begin{figure}[htbp]
\centering
\includegraphics[width=0.92\textwidth]{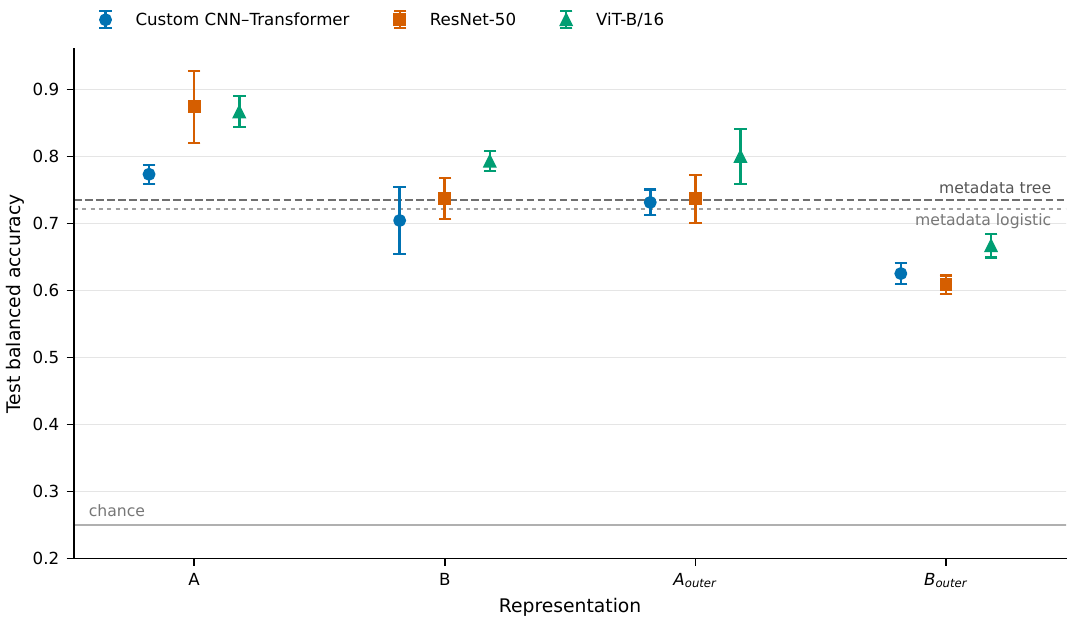}
\caption{Test balanced accuracy for each model family and image representation.
Points show the mean and bars show sample SD over the three fixed seeds. Dashed
lines are the two single-fit metadata-only controls. They are reference levels, not
uncertainty intervals.}
\label{fig:testba}
\end{figure}

The two metadata-only controls also generalized: the depth-3 decision tree obtained
0.735 balanced accuracy, 0.732 macro-F1, and 0.740 accuracy; logistic regression
obtained 0.721, 0.718, and 0.726. Figure~\ref{fig:testba} places these single-fit
controls beside the neural models. Masking the central region left substantial
predictive signal. On $A_{\mathrm{outer}}$, balanced accuracy was 0.731--0.800;
even $B_{\mathrm{outer}}$ remained at 0.608--0.667, well above chance.

Table~\ref{tab:pairedA} reports the pre-specified A-view architecture
comparisons. ResNet-50 and ViT-B/16 were nearly indistinguishable: their balanced
accuracy difference was 0.007, with a 95\% cluster-bootstrap interval spanning
$-0.048$ to 0.062. The custom model had lower point estimates, most clearly in
nebula recall (mean 0.275, versus 0.765 for ResNet-50 and 0.804 for ViT-B/16).
Several custom-versus-pretrained bootstrap intervals excluded zero, but none of the
nine randomization tests rejected after the prespecified Holm correction. This is
not a contradiction: the intervals are unadjusted 95\% effect intervals for
individual contrasts, whereas Holm controls the family-wise error rate across nine
randomization tests. They answer different inferential questions. We therefore
treat the gaps as descriptive effect estimates rather than multiplicity-adjusted
evidence of architectural superiority.

\begin{table*}[t]
\centering
\small
\caption{Paired A-view architecture comparisons. Differences are first minus
second family, averaged over the three seeds. Intervals use the paired
assignment-unit bootstrap; $p$ values use paired unit-level randomization. Holm
adjustment covers the nine displayed tests.}
\label{tab:pairedA}
\resizebox{\textwidth}{!}{%
\begin{tabular}{lllrrrr}
\toprule
First & Second & Metric & Difference & 95\% interval & Raw $p$ & Holm $p$ \\
\midrule
Custom & ResNet-50 & Accuracy & $-0.091$ & $[-0.169,-0.014]$ & 0.032 & 0.191 \\
Custom & ResNet-50 & Balanced accuracy & $-0.101$ & $[-0.156,-0.043]$ & 0.016 & 0.130 \\
Custom & ResNet-50 & Macro-F1 & $-0.126$ & $[-0.201,-0.054]$ & 0.009 & 0.077 \\
Custom & ViT-B/16 & Accuracy & $-0.082$ & $[-0.174, 0.009]$ & 0.101 & 0.404 \\
Custom & ViT-B/16 & Balanced accuracy & $-0.093$ & $[-0.159,-0.029]$ & 0.057 & 0.285 \\
Custom & ViT-B/16 & Macro-F1 & $-0.118$ & $[-0.206,-0.033]$ & 0.023 & 0.159 \\
ResNet-50 & ViT-B/16 & Accuracy & $ 0.009$ & $[-0.044, 0.066]$ & 0.873 & 1.000 \\
ResNet-50 & ViT-B/16 & Balanced accuracy & $ 0.007$ & $[-0.048, 0.062]$ & 0.842 & 1.000 \\
ResNet-50 & ViT-B/16 & Macro-F1 & $ 0.008$ & $[-0.047, 0.064]$ & 0.826 & 1.000 \\
\bottomrule
\end{tabular}
}
\end{table*}

\section{Discussion}

\subsection{Generalization and architecture}

The test establishes a limited but meaningful result: all three families generalized
to catalogue-supported physical objects absent from model development. ResNet-50
and ViT-B/16 retained balanced accuracies near 0.87, while the custom model reached
0.77. Their ranking resembles development, but the corrected tests do not support a
claim that one architecture is statistically superior in this test sample.

The custom hybrid therefore did not realize its motivating expectation of an
advantage in this small-data setting. Its largest class-specific weakness was
nebula recognition, where most errors were assignments to the open-cluster class.
The experiment provides no evidence that this evaluated hybrid system is preferable
to the two pretrained references. Because initialization and training policy differ,
it also does not establish that CNN--Transformer hybrids are intrinsically inferior.

The comparison is also deliberately practical rather than architecture-pure. The
custom model was trained from scratch, whereas ResNet-50 and ViT-B/16 received
ImageNet initialization. Their advantage combines architecture, parameterization,
optimization, and transfer learning. The near-zero ResNet--ViT differences provide
no evidence that convolution or global attention is intrinsically preferable here.

\subsection{Acquisition, catalogue metadata, and context}

The strongest methodological finding is that target identity is not the only source
of predictive information. Combined acquisition and catalogue metadata reached
balanced accuracy above 0.72. After the fixed central region was masked, the adaptive
views remained between 0.73 and 0.80, with ViT-B/16 losing only 0.067
balanced-accuracy points relative to A. These controls show that class labels covary
with acquisition choices, catalogue-derived object extent, and image content outside
the central mask. They do not establish a causal mechanism or demonstrate
classification without the target: extended nebulae and open clusters can remain
visible outside the mask, alongside neighbours, background structure, and processing
signatures.

The common-field representation reduces explicit FOV variation but is not a clean
causal intervention on FOV. Its effective $16\times16$ sampling blurs compact
targets, can truncate extended structure, and is then repeated to $224\times224$.
Consequently, A--B differences combine angular context, scale normalization, and
information loss. The B results should not be read as the cost of removing an
acquisition shortcut alone.

\subsection{Scope and uncertainty}

The 73 test objects comprise only 61 independent assignment units after overlapping
fields are grouped. This limits power, helps explain the wide intervals, and makes
the non-rejections evidence of uncertainty rather than equivalence. The three seeds
measure variability of the fixed training runs; they are not independent samples
from a model population.

The four catalogue classes are heterogeneous. DSS2 photographic bands are not
calibrated multispectral measurements, and independent band stretches suppress
absolute photometric ratios. Results apply to this acquisition and preprocessing
regime. More diverse surveys and independently collected objects are needed to test
whether the observed shortcuts persist under site, instrument, and epoch shifts.

\section{Reproducibility and data governance}

Each phase records data provenance, lineage, split membership, configurations,
checkpoints, and execution records. Independent reconstruction of all 1,480 tensors
(370 objects times four variants) verified identical numerical representations from
the source FITS products. Model selection and control training used no test tensors
or labels.

Test inference was performed once with a label-isolated procedure, and all 38 model
evaluations recorded complete predictions before the scoring process accessed test
labels. No test result triggered retraining, model replacement, reselection, or
repeated inference. During infrastructure diagnosis, one potentially inferable test
label may have been exposed to an operator after the models, protocols, and analyses
had been fixed; no prediction yet existed and no scientific artefact was modified.
We retain this event as a documented protocol deviation.

\subsection*{Data and code availability}

The curated AstroClassifier reproducibility archive for this study is publicly
available at \url{https://github.com/epmarinho/astroclassifier-repro}. It contains
the source code needed to reproduce the analyses, protocols, manifests,
configurations, aggregate statistics, and prediction and scoring records. A
versioned FITS and tensor dataset release is available from the same repository,
with checksums, provenance, and DSS use and attribution terms. Retrieval manifests
support independent reconstruction from the source archives. Version 1.0.0 of the
software and reproducibility archive is preserved at
\url{https://doi.org/10.5281/zenodo.23815956}.

\section{Conclusion}

Object-independent splitting changes what can be claimed from a small astronomical
image dataset. In \dataset{}, three neural families generalized to 73 unseen
physical objects, with adaptive-view balanced accuracy from 0.773 to 0.874. The two
pretrained references had higher point estimates than the custom CNN--Transformer,
but no primary contrast survived Holm correction, and pretraining remains inseparable
from architecture in this comparison.

The controls change the scientific interpretation more than the architecture
ranking. Combined acquisition-and-catalogue metadata and central-region-masked
images retained substantial predictive accuracy, showing that acquisition settings,
catalogue-derived properties, and contextual image content are informative even
under object-independent testing. Reliable small-data evaluation therefore requires
physical-object split units, explicit shortcut controls, and pre-specified one-pass
testing. High accuracy
on unseen objects is evidence of generalization within the collection; it is not, by
itself, evidence that a model learned the intended astronomical morphology.

\section*{Declaration of generative AI and AI-assisted technologies in the
manuscript preparation process}

The original AstroClassifier was independently developed by the author. Subsequent
improvements to the original implementation, extensive experimental testing, and
dataset audits were carried out with the assistance of OpenAI Codex.

\bibliographystyle{elsarticle-harv}
\begingroup
\small
\bibliography{cnn_transformer_refs}

@article{baron2019machine,
  author  = {Baron, Dalya},
  title   = {Machine Learning in Astronomy: A Practical Overview},
  journal = {arXiv e-prints},
  pages   = {arXiv:1904.07248},
  year    = {2019},
  url     = {https://arxiv.org/abs/1904.07248}
}

@article{dieleman2015,
  author  = {Dieleman, Sander and Willett, Kyle W. and Dambre, Joni},
  title   = {Rotation-invariant convolutional neural networks for galaxy morphology prediction},
  journal = {Monthly Notices of the Royal Astronomical Society},
  volume  = {450},
  number  = {2},
  pages   = {1441--1459},
  year    = {2015},
  doi     = {10.1093/mnras/stv632}
}

@article{huertas2018,
  author  = {Huertas-Company, Marc and Lanusse, Fran\c{c}ois},
  title   = {Deep Learning for Galaxy Morphology Classification},
  journal = {Galaxies},
  volume  = {6},
  number  = {1},
  pages   = {1},
  year    = {2018},
  doi     = {10.3390/galaxies6010001}
}

@article{harris1996,
  author  = {Harris, William E.},
  title   = {A Catalog of Parameters for Globular Clusters in the Milky Way},
  journal = {The Astronomical Journal},
  volume  = {112},
  number  = {4},
  pages   = {1487--1488},
  year    = {1996},
  doi     = {10.1086/118116},
  note    = {December 2010 catalogue revision used via HEASARC}
}

@article{fernique2015,
  author  = {Fernique, Pierre and Allen, Mark G. and Boch, Thomas and Oberto,
             Anais and Pineau, Fran\c{c}ois-Xavier and Durand, Daniel and Bot,
             Caroline and Cambr\'{e}sy, Laurent and Derriere, S\'{e}bastien and
             Genova, Fran\c{c}oise and Bonnarel, Fran\c{c}ois},
  title   = {Hierarchical Progressive Surveys: Multi-resolution {HEALPix} Data
             Structures for Astronomical Images, Catalogues, and
             3-dimensional Data Cubes},
  journal = {Astronomy and Astrophysics},
  volume  = {578},
  pages   = {A114},
  year    = {2015},
  doi     = {10.1051/0004-6361/201526075}
}

@inproceedings{boch2020hips2fits,
  author    = {Boch, Thomas and Fernique, Pierre and Bonnarel, Fran\c{c}ois and
               {Chaitra} and Bot, Caroline and Pineau, Fran\c{c}ois-Xavier and
               Baumann, Matthieu and Michel, Laurent},
  title     = {{HiPS2FITS}: Fast Generation of {FITS} Cutouts from {HiPS} Image Datasets},
  booktitle = {Astronomical Data Analysis Software and Systems XXIX},
  series    = {Astronomical Society of the Pacific Conference Series},
  volume    = {527},
  pages     = {121--124},
  year      = {2020}
}

@inproceedings{he2016deep,
  author    = {He, Kaiming and Zhang, Xiangyu and Ren, Shaoqing and Sun, Jian},
  title     = {Deep Residual Learning for Image Recognition},
  booktitle = {2016 IEEE Conference on Computer Vision and Pattern Recognition},
  pages     = {770--778},
  year      = {2016},
  doi       = {10.1109/CVPR.2016.90}
}

@inproceedings{vaswani2017,
  author    = {Vaswani, Ashish and Shazeer, Noam and Parmar, Niki and Uszkoreit,
               Jakob and Jones, Llion and Gomez, Aidan N. and Kaiser, \L{}ukasz
               and Polosukhin, Illia},
  title     = {Attention Is All You Need},
  booktitle = {Advances in Neural Information Processing Systems},
  volume    = {30},
  pages     = {5998--6008},
  year      = {2017},
  publisher = {Curran Associates, Inc.}
}

@inproceedings{dosovitskiy2020,
  author    = {Dosovitskiy, Alexey and Beyer, Lucas and Kolesnikov, Alexander and
               Weissenborn, Dirk and Zhai, Xiaohua and Unterthiner, Thomas and
               Dehghani, Mostafa and Minderer, Matthias and Heigold, Georg and
               Gelly, Sylvain and Uszkoreit, Jakob and Houlsby, Neil},
  title     = {An Image Is Worth 16x16 Words: Transformers for Image Recognition at Scale},
  booktitle = {International Conference on Learning Representations},
  pages     = {1--22},
  year      = {2021},
  url       = {https://openreview.net/forum?id=YicbFdNTTy}
}

@article{khan2021,
  author    = {Khan, Salman and Naseer, Muzammal and Hayat, Munawar and Zamir,
               Syed Waqas and Khan, Fahad Shahbaz and Shah, Mubarak},
  title     = {Transformers in Vision: A Survey},
  journal   = {ACM Computing Surveys},
  volume    = {54},
  number    = {10s},
  pages     = {1--41},
  articleno = {200},
  year      = {2022},
  doi       = {10.1145/3505244}
}

@article{shorten2019,
  author  = {Shorten, Connor and Khoshgoftaar, Taghi M.},
  title   = {A Survey on Image Data Augmentation for Deep Learning},
  journal = {Journal of Big Data},
  volume  = {6},
  number  = {1},
  pages   = {60},
  year    = {2019},
  doi     = {10.1186/s40537-019-0197-0}
}

@article{geirhos2020shortcut,
  author  = {Geirhos, Robert and Jacobsen, J{\"o}rn-Henrik and Michaelis, Claudio
             and Zemel, Richard and Brendel, Wieland and Bethge, Matthias and
             Wichmann, Felix A.},
  title   = {Shortcut Learning in Deep Neural Networks},
  journal = {Nature Machine Intelligence},
  volume  = {2},
  pages   = {665--673},
  year    = {2020},
  doi     = {10.1038/s42256-020-00257-z}
}

@article{kapoor2023leakage,
  author  = {Kapoor, Sayash and Narayanan, Arvind},
  title   = {Leakage and the Reproducibility Crisis in Machine-Learning-Based Science},
  journal = {Patterns},
  volume  = {4},
  number  = {9},
  pages   = {100804},
  year    = {2023},
  doi     = {10.1016/j.patter.2023.100804}
}

@article{wenger2000simbad,
  author  = {Wenger, M. and Ochsenbein, F. and Egret, D. and Dubois, P. and
             Bonnarel, F. and Borde, S. and Genova, F. and Jasniewicz, G. and
             Lalo{\"e}, S. and Lesteven, S. and Monier, R.},
  title   = {The {SIMBAD} Astronomical Database: The {CDS} Reference Database for Astronomical Objects},
  journal = {Astronomy and Astrophysics Supplement Series},
  volume  = {143},
  number  = {1},
  pages   = {9--22},
  year    = {2000},
  doi     = {10.1051/aas:2000332}
}
\endgroup
\end{document}